\documentstyle[preprint,prc,aps]{revtex}

\begin{document}

%
%
%
%
\catcode `@=11    
%
\newenvironment{symbols}{\ifsymboloutput\symch@ck\fi}{\ifsymboloutput%
                                                       \newpage               
   \immediate\closeout\symbolwrite \symbolopenfalse
   {%
   \pagenumbering{roman}
   \Large\centerline{Symbols}}\bigskip
   \input symbols.aux
   \newpage
   \pagenumbering{arabic}
   \protect\setcounter{page}{1}
   \fi}
\newif\ifsymboloutput    
\newif\ifsymbolopen      
\newwrite\symbolwrite
\def\symch@ck{\ifsymbolopen\else        
   \immediate\openout\symbolwrite=symbols.aux
   \symbolopentrue\fi}
\def\symdef#1#2{\gdef#1{#2}
      \ifsymboloutput\symch@ck%
         \immediate\write\symbolwrite{$$ \hbox{\string\verb+\string#1+}
                \noexpand\qquad\noexpand\longrightarrow\noexpand\qquad
                           \string#1 $$}%
      \fi}
\def\symbolout{\newpage               
   \immediate\closeout\symbolwrite \symbolopenfalse
   {%
   \pagenumbering{roman}
   \Large\centerline{Symbols}}\bigskip
   \input symbols.aux
   \newpage
   \pagenumbering{arabic}
   \protect\setcounter{page}{1}
   }
\catcode `@=12    
%
%
\symboloutputtrue      
%
%
%
%
\begin{symbols}    
  \symdef\e{{\rm e}}    
  \symdef\Eq{E_q}                    
  \symdef\Eqbar{\overline E_q}
  \symdef\Eqstar{E_q^{\ast}}
  \symdef\kfermi{k_{{\scriptscriptstyle\rm F}}}  
  \symdef\MN{M_{\scriptscriptstyle\rm N}}      
  \symdef\Mstar{M^\ast_{\scriptscriptstyle\rm N}}    
  \symdef\PiN{\Pi_{{\scriptscriptstyle\rm N}}}   
  \symdef\Piq{\Pi_q}
  \symdef\Piqtilde{\widetilde\Pi_q}
  \symdef\Pis{\Pi_s}
  \symdef\Pistilde{\widetilde\Pi_s}
  \symdef\Piu{\Pi_u}
  \symdef\Piutilde{\widetilde\Pi_u}
  \symdef\psibar{\overline\psi}        
  \symdef\qbarq{\langle\overline q q \rangle_{\rho}}  
  \symdef\qdaggerq{\langle q^\dagger q \rangle_{\rho}}  
  \symdef\qdotu{q\cdot u}    
  \symdef\qslash{\rlap{/}{q}}       
  \symdef\qtildeslash{\rlap{/}{\widetilde q}}
  \symdef\qvec{{\bf q}}        
  \symdef\residue{\lambda^2}               
  \symdef\residuebar{\overline\lambda^2}
  \symdef\rhoph{\rho^{{\rm ph}}}    
  \symdef\rhoth{\rho^{{\rm th}}}    
  \symdef\rhoN{\rho_{{\scriptscriptstyle\rm N}}}  
  \symdef\sigmav{\Sigma_{{\rm v}}}    
  \symdef\uslash{\rlap{/}{u}}        
  \symdef\utildeslash{\rlap{/}{\widetilde u}}
  \symdef\wzero{\omega_0}              
  \symdef\wzerobar{\overline\omega_0}  
\end{symbols}
%
%

\draft

\preprint{OSU--0901}

\title{Spectral asymmetries in nucleon sum rules\\
        at finite density}

\author{R. J. Furnstahl}
\address{Department of Physics \\
         The Ohio State University,\ \ Columbus, Ohio\ \ 43210}
%

%
\date{October, 1993}
\maketitle

\begin{abstract}
Apparent inconsistencies between different formulations of nucleon sum
rules at finite density
are resolved through a proper accounting of asymmetries in the spectral
functions between positive- and negative-energy states.
\end{abstract}


\narrowtext

Relativistic hadronic models
suggest that nucleons propagating in nuclear matter
develop large, attractive scalar and repulsive vector
self-energies \cite{SEROT92}.
Evidence
supporting this picture in finite-density quantum chromodynamics (QCD) has
emerged from some recent QCD sum rule
calculations \cite{DRUKAREV91,HATSUDA91,lp1,HENLEY93}.
In particular, a rough
connection is established between the nucleon scalar self-energy $\Mstar$
and the scalar
condensate $\qbarq$ in medium, and between the nucleon vector self-energy
$\sigmav$ and the
vector condensate $\qdaggerq$.
An alternative
formulation of the sum rules \cite{tokyo}, however,
apparently gives qualitatively different  results, in which
the vector self-energy takes the opposite sign, so that
the vector condensate induces attraction rather
than repulsion.
In this note, we resolve this  ambiguity by properly accounting
for spectral asymmetries in the
finite-density sum rules for the nucleon, showing that the two formulations
are each consistent with relativistic phenomenology.

The QCD sum rules for a nucleon in nuclear matter focus on a correlator
of interpolating fields for the nucleon:
\begin{equation}
	\PiN(q) = i \int\! d^4x \, \e^{iq\cdot x}\,
	         \langle \Phi_0 | T\{\eta(x) \overline\eta(0)\} |
                         \Phi_0  \rangle \ ,
	\label{eq:correlator}
\end{equation}
where $|\Phi_0\rangle$ is the nuclear matter ground state,
which is characterized
by four-velocity $u^\mu$ and baryon density $\rhoN$ in the rest frame.
The interpolating field $\eta(x)$ is taken here
and in Refs.~\cite{lp1,tokyo} to be that
advocated by Ioffe \cite{IOFFE81}.
(See Ref.~\cite{lp3} for a treatment of more general
interpolating fields and for explicit expressions.)

Because two four-vectors are available,
the correlator can be decomposed into invariant functions in
two  ways:\footnote{Relativistic covariance, time reversal, and parity
imply that there are only three independent invariant functions \cite{lp1}.}
\begin{eqnarray}
	\PiN(q) & = & \qslash\, \Piq(q^2,\qdotu)
                                + \Pis(q^2,\qdotu)  \nonumber \\
            &   &  \null    + \uslash\, \Piu(q^2,\qdotu)
                                  	\label{eq:ourway} \\
	 & = &  \qtildeslash\, \Piqtilde(q^2,\qdotu)
	           + \Pistilde(q^2,\qdotu)   \nonumber \\
             &   &  \null   +  \uslash\, \Piutilde(q^2,\qdotu)    \ ,
	\label{eq:theirway}
\end{eqnarray}
where $\tilde q^\mu \equiv q^\mu - (\qdotu) u^\mu$.
The two sets of invariant functions are trivially related:
$\Piqtilde = \Piq$,
$\Pistilde = \Pis$, and $\Piutilde = (\qdotu) \Piq + \Piu$.
These two alternative decompositions  will lead to the two sets of sum
rules we consider.

The analytic structure of the correlator $\PiN$, and consequently
the invariant
functions in Eqs.~(\ref{eq:ourway}) and (\ref{eq:theirway}),
is revealed by a
standard Lehmann representation in energy $\omega$, at fixed
three-momentum $\qvec$ \cite{FETTER71,lp1}.
[We work for {\it convenience\/} in the rest frame of
the matter, where $u^\mu = (1,{\bf 0}$) and $q^\mu = (\omega,\qvec)$.]
The invariant functions are  found to have singularities only on the
real axis.
In the upper half plane, $\PiN(\omega,\qvec)$
is equal to the retarded correlator and
in the lower half plane to the advanced correlator.
We will be concerned only with the discontinuities across the real axis
(that is, the spectral functions),
and therefore the infinitesimals that differentiate retarded, advanced,
and time-ordered correlators for real $\omega$ are not relevant here.

We exploit the analytic structure of the correlator by considering
integrals over contours
running above and below the real axis, and then closing in the
upper and lower half planes, respectively (see Ref.~\cite{lp1}).
By approximating the correlator in the different regions of integration
and applying Cauchy's theorem,
we can derive a general class of  sum rules, which
manifest the  duality between the physical hadronic spectrum and
the spectral function calculated in a QCD expansion \cite{KRASNIKOV83}:
\begin{equation}
	\int_{-\wzerobar}^{\wzero} W(\omega)
                            \rhoph(\omega,\qvec)\,  d\omega
	-
	\int_{-\wzerobar}^{\wzero} W(\omega)
                            \rhoth(\omega,\qvec)\,  d\omega  = 0 \ .
	\label{eq:fesr}
\end{equation}
Here $W(\omega)$ is a smooth (analytic) weighting function and the
spectral densities $\rhoph$ and $\rhoth$ are proportional to the
discontinuities of the invariant functions across the real axis.
(These sum rules can also be derived by expanding dispersion relations for
retarded and advanced correlators with external frequency $\omega'$
in the limit $\omega' \rightarrow i\infty$.)
The phenomenological spectral density $\rhoph$ models the low-energy physical
spectrum, while the theoretical spectral density $\rhoth$ is
approximated using
an operator product expansion (OPE) applied to $\PiN$.
The QCD sum rule approach assumes that,
with suitable choices for $W$ and the effective continuum thresholds
$\wzero$ and $-\wzerobar$, each integral can be reliably calculated and
meaningful results extracted.

The thresholds $\wzero$ and $-\wzerobar$ act as effective boundaries
beyond which the physical spectrum $\rhoph$, when moderately smoothed,
is identical to that of the leading contributions to $\rhoth$.
At zero density, the discrete space-time
symmetries imply that $\rho(-\omega,\qvec)
= -\rho(+\omega,\qvec)$, and so we are led to take $\wzerobar = \wzero$.
That is, charge conjugation implies that the free-space spectral
function corresponding to positive-energy states (like the nucleon) is
equal in magnitude to that of the corresponding negative-energy states
(like the antinucleon).
Since the relative sign of the positive- and negative-energy spectral
functions is negative,
only a weighting function odd in $\omega$ yields a nonzero result
in Eq.~(\ref{eq:fesr}).
In this case,
we can convert the integral to one over $\omega^2$ and then
over $s = \omega^2 - \qvec^2$.
The end result features only manifestly covariant integrals, as one would
expect.

Useful choices for the weighting function at zero density are
$W(\omega) = \omega \e^{-\omega^2/M^2}$,
which generates the conventional Borel sum rules \cite{SHIFMAN79,lp1},
or polynomials in $\omega$, which generate the so-called
finite energy sum rules (FESR) \cite{KRASNIKOV83}.
(Actually, $\omega$ times a monomial in $s$ is used in the
latter case, since
$\rho = \rho(s)$ in vacuum.)

At finite density, ordinary charge conjugation symmetry is broken by
the nonzero baryon number of the medium.
This is clear from physical considerations:
the propagation of a nucleon  in ordinary nuclear matter
is quite different from the propagation of an antinucleon.
The consequence for the spectral function $\rhoph(\omega,\qvec)$
is an asymmetry with
respect to positive and negative $\omega$.
Asymmetries also appear
naturally in $\rhoth(\omega,\qvec)$ by generalizing the
OPE to finite density [see below].
As a result, we are also compelled to adopt an asymmetric duality interval
so that $\wzero \neq \wzerobar$; this is
the feature missed in Ref.~\cite{tokyo}.

Following Ref.~\cite{lp1},
we saturate the phenomenological integral in Eq.~(\ref{eq:fesr})
with a  quasiparticle
pole ansatz:
\begin{equation}
	\PiN \propto {1 \over \qslash - \uslash \sigmav - \Mstar}  \ .
                  	\label{eq:ansatz}
\end{equation}
This implies
\begin{eqnarray}
	\rhoph_q & = & \widetilde\rhoph_q \nonumber \\
      & = &
                   {\pi\over 2\Eqstar}
               \left[ \residue\delta(\omega-\Eq) -
	                   \residuebar\delta(\omega-\Eqbar) \right]  \ ,
                             	\label{eq:rhoq} \\
	\rhoph_s & = & \widetilde\rhoph_s \nonumber  \\
      & = &
                    {\pi\over 2\Eqstar}  \Mstar
                 \left[ \residue\delta(\omega-\Eq) -
	                   \residuebar\delta(\omega-\Eqbar) \right]  \ ,
                            	\label{eq:rhos} \\
	\rhoph_u & = &  -{\pi\over 2\Eqstar} \sigmav
                          \left[ \residue\delta(\omega-\Eq) -
	                   \residuebar\delta(\omega-\Eqbar) \right]  \ ,
                         	\label{eq:rhou}  \\
                  \widetilde\rhoph_u & = &
                              {\pi\over 2\Eqstar}  \Eqstar
                              \left[ \residue\delta(\omega-\Eq) +
	                   \residuebar\delta(\omega-\Eqbar) \right]  \ ,
                                \label{eq:rhoutilde}
\end{eqnarray}
where $\Eqstar \equiv (q^2 + \Mstar{}^2)^{1/2}$,
$\Eq = \sigmav + \Eqstar$, and $\Eqbar = \sigmav - \Eqstar < 0$.
The notation is that $\rho_i$ ($\widetilde\rho_i$) corresponds to
the discontinuity of $\Pi_i$ ($\widetilde\Pi_i$), with $i = \{q,s,u\}$.
We have allowed for different residues $\residue$ and $\residuebar$
 for the positive and negative-energy  poles, respectively.
Different residues would naturally arise, for example,
from the energy dependence of the
self-energies.
The form of the ansatz in Eq.~(\ref{eq:ansatz}) is otherwise
constrained by Lorentz
covariance.
While a sharp quasinucleon state
represents an extreme simplification of the actual
spectrum, it is not unrealistic when smeared over
energy scales of several hundred MeV, as we do in the sum rules.
The physics motivation for the ansatz is discussed further in Ref.~\cite{lp1}.

The operator product expansion for $\PiN$ is developed in
Refs.~\cite{lp1,lp2,lp3}.
We keep only the terms from the leading diagrams here (corresponding to those
considered in the first paper of Ref.~\cite{tokyo}) to illustrate our point.
For a quantitative analysis, one would include at least the contributions
through terms involving four-quark condensates.
The expansion is (suppressing parts that do not contribute to $\rhoth$)
\begin{eqnarray}
   \PiN(q) & = &
         \qslash \Bigl[  -{1 \over 64\pi^4} (q^2)^2\ln(-q^2)
             \nonumber \\
     &   & \quad \null
               + {1\over 3\pi^2} \qdotu \ln(-q^2) \qdaggerq \Bigr]
        \nonumber\\   &   &
         + {1\over 4\pi^2} q^2 \ln(-q^2) \qbarq
       + \uslash  {2\over 3\pi^2} q^2 \ln(-q^2) \qdaggerq  \ ,
          \nonumber \\  &   &  \null
                                 \label{eq:ope}
\end{eqnarray}
which implies (specializing to the rest frame again)
\begin{eqnarray}
    \rhoth_q & = & \widetilde\rhoth_q =
    \pi [ \theta(\omega-|\qvec|) - \theta(-\omega-|\qvec|) ]  \nonumber\\
   &   &  \quad\qquad\null\times
   \left[ {1\over 64\pi^4} (q^2)^2  - {1\over 3\pi^2}\omega \qdaggerq \right]
         \ ,
                                \label{eq:rhothq}  \\
  \rhoth_s & = & \widetilde\rhoth_s =
         -\pi [ \theta(\omega-|\qvec|) - \theta(-\omega-|\qvec|) ] \nonumber\\
   &   &  \quad\qquad\null\times
      \left[ {1\over 4\pi^2} q^2 \qbarq \right]  \ ,
                                 \label{eq:rhoths}  \\
  \rhoth_u & = &
    -\pi [ \theta(\omega-|\qvec|) - \theta(-\omega-|\qvec|) ]
   \left[ {2\over 3\pi^2} q^2 \qdaggerq \right]  \ ,
                                  \label{eq:rhothu}  \\
   \widetilde\rhoth_u & = &
    \pi [ \theta(\omega-|\qvec|) - \theta(-\omega-|\qvec|) ]  \nonumber\\
   &   & \null\times
  \left[ {\omega\over 64\pi^4}(q^2)^2
       - ({2\over 3\pi^2}q^2 + {1\over 3\pi^2}\omega^2)\qdaggerq \right] \ .
                                  \label{eq:rhothutilde}
\end{eqnarray}
We now have all the ingredients needed to construct QCD sum rules
using Eq.~(\ref{eq:fesr}).

The phenomenological expectation from Relativistic Brueckner
Hartree-Fock (RBHF) and mean-field calculations,
as well as from Dirac phenomenology, is that at nuclear matter density
$\Mstar \approx 0.65 \MN$ and $\sigmav \approx 0.35 \MN$, and  the sum
is relatively constant with density \cite{SEROT92}.
This expectation is supported by a simple version of the sum rules
that is discussed in Ref.~\cite{lp1}.
(A more complete sum rule is, however, not yet
conclusive because of uncertainties from
higher-order condensates \cite{lp1,lp3}.)
In these rules, $\Mstar$ is correlated with $\qbarq$, and $\sigmav>0$ with
$\qdaggerq$.

The authors of
two recent preprints, however, obtain very different qualitative results.
In Ref.~\cite{tokyo}, the  decomposition in Eq.~(\ref{eq:theirway})
was (implicitly) adopted, and
applied with $\qvec =0$.  The apparent advantage is that
 a direct projection onto the positive-energy quasinucleon is possible.
That is, ${1\over2}(1 + \gamma_0)$
projects onto the positive-energy quasinucleon
and ${1\over2}(1 - \gamma_0)$ projects onto the negative-energy
quasinucleon.
{}From these projected sum rules, $\qdaggerq$ was found in Ref.~\cite{tokyo}
to be associated with
{\it  attraction\/} rather than repulsion, implying that $\sigmav < 0$.
The net result was a very large energy shift of the quasinucleon pole.
Here we point out the missing elements
in these calculations, which drastically affect the conclusions.

As noted above,
the essential point is that the asymmetry in the phenomenological spectral
density must necessarily be reflected in an asymmetry in the duality
interval; that is, $\wzerobar \neq \wzero$.
This point and its semi-quantitative implications are made clear by
considering FESR's, using Eqs.~(\ref{eq:rhothq})--(\ref{eq:rhothutilde})
with monomials in $\omega$ as the
weighting functions $W(\omega)$.

We start with the first decomposition, Eq.~(\ref{eq:ourway}).
Given a quasiparticle ansatz for the nucleon Eq.~(\ref{eq:ansatz}),
this is a natural
decomposition for isolating the scalar self-energy $\Mstar$
(by considering $\Pis/\Piq$)
and the vector self-energy $\sigmav$ (by considering $\Piu/\Piq$).
We apply Eq.~(\ref{eq:fesr}) to each of the three functions $\rho_q$,
$\rho_s$, and $\rho_u$,
first with $W(\omega)=1$
and then with $W(\omega) = \omega$.
We can derive further sum rules, of course, but these are sufficient to
make our point.
For simplicity and for a clear comparison to Ref.~\cite{tokyo},
we take $\qvec=0$
from here on,
so that $q^2 \rightarrow \omega^2$
in Eqs.~(\ref{eq:rhothq})--(\ref{eq:rhothutilde}).
We obtain
\widetext
\begin{eqnarray}
	\null & q: & \quad {1\over 2\Mstar} (\residue-\residuebar) =
	     {1\over 320\pi^4} (\wzero^5 - \wzerobar^5)
	      - {1\over 6\pi^2} \qdaggerq (\wzero^2 + \wzerobar^2)  \ ,
                          	\label{eq:q0} \\
	\null & \phantom{q:} & \quad
	 {1\over 2\Mstar} [\Mstar(\residue+\residuebar) +
	                       \sigmav (\residue-\residuebar)] =
	    {1\over 384\pi^4} (\wzero^6 + \wzerobar^6)
	      - {1\over 9\pi^2} \qdaggerq (\wzero^3 - \wzerobar^3)  \ ,
                        	\label{eq:q1} \\
	\null & s: & \quad {1\over 2} (\residue-\residuebar) =
	     -{1\over 12\pi^2} \qbarq (\wzero^3 - \wzerobar^3)  \ ,
                            	\label{eq:s0} \\
	\null & \phantom{s:} & \quad
	 {1\over 2} [\Mstar(\residue+\residuebar) +
	                       \sigmav (\residue-\residuebar)] =
	     -{1\over 16\pi^2}\qbarq (\wzero^4 + \wzerobar^4)  \ ,
                          	\label{eq:s1} \\
	\null & u: & \quad {\sigmav\over 2\Mstar} (\residue-\residuebar) =
	         {2\over 9\pi^2}\qdaggerq (\wzero^3 - \wzerobar^3)  \ ,
                           	\label{eq:u0} \\
	\null & \phantom{u:} & \quad
	 {\sigmav\over 2\Mstar} [\Mstar(\residue+\residuebar) +
	                       \sigmav (\residue-\residuebar)] =
	     {1\over 6\pi^2} \qdaggerq(\wzero^4 + \wzerobar^4)  \ .
                            	\label{eq:u1}
\end{eqnarray}
\narrowtext
One should be cautious about  quantitative results extracted from
these sum rules, but the qualitative features should persist in more
sophisticated treatments.

The  terms in the operator product expansion (OPE)
odd in $\omega$ vanish at zero density.
But at finite density, they imply, through duality, an asymmetry in the
physical spectral functions, as manifested in Eq.~(\ref{eq:q0}).
There are  many more such terms as we extend the OPE to
higher dimension.

If we expand in density, then from Eq.~(\ref{eq:q0}) we see that
(recall that $\residue = \residuebar$
and $\wzero=\wzerobar$ at $\rhoN=0$ from charge conjugation
symmetry)
\begin{eqnarray}
	\wzero - \wzerobar & \sim & O(\rhoN)  \ ,
           	\label{eq:wdiff} \\
	\residue - \residuebar & \sim & O(\rhoN)  \ .
           	\label{eq:resdiff}
\end{eqnarray}
If we drop terms of higher order in $\rhoN$,
Eqs.~(\ref{eq:q1}), (\ref{eq:s1}), and (\ref{eq:u1}) become
\begin{eqnarray}
	 {1\over 2} (\residue+\residuebar) & = &
	    {1\over 384\pi^4} (\wzero^6 + \wzerobar^6)  \ ,
                                \label{eq:q1p}  \\
	 {1\over 2} \Mstar(\residue+\residuebar)  & = &
	     -{1\over 16\pi^2}\qbarq (\wzero^4 + \wzerobar^4)  \ ,
                           	\label{eq:s1p} \\
	 {1\over 2} \sigmav (\residue+\residuebar) & = &
	     {1\over 6\pi^2} \qdaggerq(\wzero^4 + \wzerobar^4)  \ .
                             	\label{eq:u1p}
\end{eqnarray}
We can divide Eqs.~(\ref{eq:s1p}) and (\ref{eq:u1p}) by (\ref{eq:q1p}) to find
\begin{equation}
	\Mstar = -{24\pi^2\over \langle \wzero^2 \rangle} \qbarq
	  \qquad \mbox{and}\qquad
	\sigmav = {64\pi^2\over \langle \wzero^2 \rangle} \qdaggerq  \ ,
                            	\label{eq:ioffe}
\end{equation}
where $\langle \wzero^2 \rangle \approx {1\over2} (\wzero^2 + \wzerobar^2)$
to this order.
Plugging in a typical continuum threshold \cite{lp1,lp3},
we find roughly the
magnitudes, density dependence, and scalar-vector cancellations
expected for the self-energies from relativistic phenomenology.
Note that $\sigmav$ is unambiguously positive.
The scale of $\wzero-\wzerobar$ is set by Eq.~(\ref{eq:q0}), which in
conjunction with Eq.~(\ref{eq:s0}) and the results for $\sigmav$ and $\Mstar$
in Eq.~(\ref{eq:ioffe}),
implies that $\wzero-\wzerobar \approx \sigmav$.

Now we consider the second decomposition,
Eq.~(\ref{eq:theirway}), and extract the analogous sum rules.
The rules from $\Piqtilde$ and $\Pistilde$ are the same as above.
The only different sum rules come from $\Piutilde$:
\widetext
\begin{eqnarray}
	\null & \widetilde u: & \quad  {1\over 2}(\residue + \residuebar) =
	     {1\over 384\pi^4} (\wzero^6 + \wzerobar^6)
	        - {\qdaggerq \over 3\pi^2} (\wzero^3 - \wzerobar^3) \ ,
                           	\label{eq:utilde0} \\
	\null & \phantom{u:} & \quad {\sigmav\over 2}(\residue + \residuebar)
	    + {\Mstar\over 2}(\residue - \residuebar) =
	    -{\qdaggerq\over 4\pi^2}(\wzero^4 + \wzerobar^4)
	     + {1\over 448\pi^4} (\wzero^7 - \wzerobar^7) \ .
                            	\label{eq:utilde1}
\end{eqnarray}
The underlying
problem with Ref.~\cite{tokyo} is made clear by Eq.~(\ref{eq:utilde1}).
If $\wzerobar=\wzero$ is assumed [and therefore $\residuebar=\residue$
from Eq.~(\ref{eq:s0})], one
concludes  that $\sigmav < 0$.

Adding the sum rules for $\Pistilde = \Pis$ and $\Piutilde$ (that is, making
the ${1\over2}(1+\gamma_0)$ projection), we obtain:
\begin{eqnarray}
	\residue & = &
               {1\over 384\pi^4} (\wzero^6 + \wzerobar^6)
               - {\qdaggerq \over 3\pi^2} (\wzero^3 - \wzerobar^3)
               - {\qbarq \over 12\pi^2} (\wzero^3 - \wzerobar^3) \ ,
                            	\label{eq:proj0} \\
	\residue (\sigmav + \Mstar) & = &
	        - {\qbarq \over 16\pi^2} (\wzero^4 + \wzerobar^4)
	        - {\qdaggerq \over 4\pi^2} (\wzero^4 + \wzerobar^4)
	     + {1\over 448\pi^4} (\wzero^7 - \wzerobar^7) \ .
                            	\label{eq:proj1}
\end{eqnarray}
Dividing these equations gives a sum rule for
the positive-energy pole:
\begin{eqnarray}
     \Eq(\qvec=0) & = & \sigmav + \Mstar  \nonumber \\
	   & = &    { - (\qbarq + 4\qdaggerq) (\wzero^4 + \wzerobar^4)
	     + {\textstyle 1\over \textstyle 28\pi^2}
                  (\wzero^7 - \wzerobar^7)
          \over
  	      {\textstyle 4\over \textstyle 3}
           \Bigl[{\textstyle 1\over \textstyle 32\pi^2}
                   (\wzero^6 + \wzerobar^6)
               -  (\qbarq + 4\qdaggerq)(\wzero^3 - \wzerobar^3)\Bigr]
                       }   \ .
                                 \label{eq:ratio}
\end{eqnarray}
\narrowtext
This is the FESR analog of the Borel sum-rule
equations derived in Ref.~\cite{tokyo}.
Again, if $\wzerobar=\wzero$ is assumed, $\qdaggerq$ and the
change with density of $\qbarq$ both apparently reduce $\Eq$ from its
zero density value of $\MN$.
However, once $\wzerobar \neq \wzero$ is allowed, several
new terms contribute.
Numerical estimates using  $\wzero-\wzerobar = \sigmav$ from
Eq.~(\ref{eq:ioffe}),
condensate values from Ref.~\cite{lp1},
and $\wzero \approx 1.6\mbox{--}1.7\,$GeV
show that $\Eq(\qvec=0)$ is essentially
constant with density rather than decreasing precipitously, as found
in Ref.~\cite{tokyo}.\footnote{These rather high values for $\wzero$ are
needed in the FESR at $\rhoN = 0$ to obtain a sensible result for $\MN$.
In addition, we note that the sum rules imply that $\wzero + \wzerobar$
is roughly constant with density.}

This quantitative result, which involves various cancellations,
is very sensitive to details of the calculation:
the choice of weighting function, the continuum threshold, the number
of terms kept in the OPE.
In contrast, sum rules derived from the first decomposition of $\PiN$,
which lead to separate expressions for the scalar and vector self-energies
rather than for the sum, are more robust.
In Refs.~\cite{lp1,lp3},
the  decomposition of Eq.~(\ref{eq:ourway}) was considered,
with an asymmetric weighting function used to
suppress contributions from the negative-energy quasinucleon in favor
of the positive-energy quasinucleon.
The possibility $\wzero \neq \wzerobar$ was not considered in these
references; however,
explicit calculations show that making this generalization is numerically
unimportant.
Thus there is a qualitative difference in  the sensitivity of the
two correlator decompositions [Eqs.~(\ref{eq:ourway}),(\ref{eq:theirway})]
to the (uncertain) details of
the continuum contribution.

In summary, we have reexamined the
results of Ref.~\cite{tokyo}, which implied that certain QCD sum rules
predict that the vector self-energy
of a nucleon in nuclear matter is attractive, rather than repulsive as
implied by relativistic phenomenology and other sum rules \cite{lp1}.
By properly accounting for asymmetries in the spectral functions
between positive- and negative-energy states, we have shown that there
is no inconsistency, and that all formulations lead to a repulsive vector
interaction.

\acknowledgments

I am grateful to Tetsuo Hatsuda for suggesting that I consider
the FESR,  and for useful discussions.
I thank D. K. Griegel, E. M. Henley, S. H. Lee, and B. D. Serot for useful
comments.
This work was supported in part by the National Science Foundation
under Grants No.\ PHY--9203145 and PHY--9258270, and the Sloan
Foundation.


\begin{references}
%
\bibitem{SEROT92}B. D. Serot, Rep.\ Prog.\ Phys.\ {\bf 55}, 1855 (1993).
%
\bibitem{DRUKAREV91}E. G. Drukarev and E. M. Levin, Prog.\ Part.\
          Nucl.\ Phys.\ {\bf 27}, 77 (1991).
%
\bibitem{HATSUDA91}T. Hatsuda, H. H{\o}gaasen, and M. Prakash,
          Phys.\ Rev.\ C {\bf 42}, 2212 (1990);
          Phys.\ Rev.\ Lett.\ {\bf 66}, 2851 (1991).
%
\bibitem{lp1}R.~J. Furnstahl, D.~K. Griegel, and T.~D. Cohen,
        Phys.\ Rev.\ C {\bf 46}, 1507 (1992).
%
\bibitem{HENLEY93}E. M. Henley, and J. Pasupathy, Nucl.\ Phys.\
         {\bf A556}, 467 (1993).
%
\bibitem{tokyo}Y. Kondo and O. Morimatsu, Institute for Nuclear Study
   report INS-Rep.-933, (June 1992) and INS-Rep.-965, (June 1993).
%
\bibitem{IOFFE81}B. L. Ioffe, Nucl.\ Phys.\ {\bf B188}, 317 (1981);
            {\bf B191}, 591 (E) (1981).
%
\bibitem{lp3}X. Jin, M. Nielsen, T.~D. Cohen, R.~J. Furnstahl,
      and D.~K. Griegel,
      Phys.\ Rev.\ C (in press).
%
\bibitem{FETTER71}A. L. Fetter and J. D. Walecka, {\it Quantum Theory of
                  Many-Particle Systems} (McGraw-Hill, New York, 1971).

\bibitem{KRASNIKOV83}N. V. Krasnikov, Z. Phys.\ C {\bf 19}, 301 (1983).
%
\bibitem{SHIFMAN79}M.~A.\ Shifman, A.~I.\ Vainshtein, and V.~I.\ Zakharov,
                     Nucl.\ Phys.\ {\bf B147}, 385 (1979); \newline
              L.~J.\ Reinders, H.~Rubinstein and S.~Yazaki,
                Phys.\ Rep.\ {\bf 127}, 1 (1985), and references therein.
%
\bibitem{lp2}X. Jin,  T.~D. Cohen, R.~J. Furnstahl,
     and D.~K. Griegel, Phys.\ Rev.\ C {\bf 47}, 2882 (1993).

\end{references}
\end{document}